# Artificial Intelligence Satellite Telecommunication Testbed using Commercial Off-The-Shelf Chipsets

Luis M. Garcés-Socarrás[*,1], Amirhossein Nik[1], Flor Ortiz[1], Juan A. Vásquez-Peralvo[1], Jorge L. González-Rios[1], Mouhamad Chehailty[1], Marcele Kuhfuss[1], Eva Lagunas[1], Jan Thoemel[1], Sumit Kumar[1], Vishal Singh[1], Juan C. Merlano Duncan[1], Sahar Malmir[1], Swetha Varadajulu[1], Jorge Querol[1], Symeon Chatzinotas[1]

[1]*Interdisciplinary Centre for Security, Reliability and Trust (SnT), University of Luxembourg, Luxembourg*

**The Artificial Intelligence Satellite Telecommunications Testbed (AISTT), part of the ESA project SPAICE, is focused on the transformation of the satellite payload by using artificial intelligence (AI) and machine learning (ML) methodologies over available commercial off-the-shelf (COTS) AI-capable chips for onboard processing. The objectives include validating artificial intelligence-driven SATCOM scenarios such as interference detection, spectrum sharing, radio resource management, decoding, and beamforming. The study highlights hardware selection and payload architecture. Preliminary results show that ML models significantly improve signal quality, spectral efficiency, and throughput compared to conventional payload. Moreover, the testbed aims to evaluate the performance and the use of AI-capable COTS chips in onboard SATCOM contexts.**

## 1 Introduction

Artificial intelligence (AI) is experiencing significant advancements, spreading throughout our daily lives to the most complex and technical domains. The benefits of automated tasks, enhanced efficiency, increased precision, and innovation, coupled with predictive analysis features, can significantly improve the effectiveness of traditional methods.

AI, especially machine learning (ML), is used in space for navigation, surveillance, remote sensing, and satellite communication, among others [1]. In the specific area of satellite communications, AI can solve complex tasks in a reasonable time, while traditional optimization methods are computationally expensive. This capability is important in the highly dynamic environment of low-earth orbit (LEO) satellites.

Very high-throughput satellites (VHTS) face inconsistent coverage, leading to a deficit in the necessary capacity in certain beams that do not meet the traffic demand while exceeding it in others [2, 3]. This issue can be addressed by integrating ML with other methods, such as flexible payload, and adaptive beamforming, which facilitate the allocation of payload resources according to the capacity requirements [4].

Although AI can be executed on several devices, the processor architecture determines the inference's performance. Current onboard processing uses space-certified general purpose processors (GPPs), costly application-specific integrated circuits (ASICs), or field-programmable gate arrays (FPGAs). Space-certified components, such as radiation hardness by design (RADHAR) or radiation tolerant (RT) devices, are specifically designed to overcome radiation. They undergo a highly stringent certification process that spans several years, leading to the use of legacy components [1, 5–7]. The increased performance demands of onboard processors to satisfy the accelerated data rates and autonomy requirements have rendered current space-graded processors obsolete [8].

Exploring non-qualified commercial off-the-shelf (COTS) devices in space offers access to advanced features despite radiation challenges [5, 8–10]. In light of the anticipated mass production of LEO constellations in the future, there has been a surge of interest in the research into COTS chipsets [1]. This is due to their potential to lower the cost of SATCOM processing, their availability on the market, and their ability to shorten the development time.

The Graphic Processing Unit (GPU) for Space (GPU4S) project [11, 12] is an example of the interest of AI-capable COTS devices for onboard space applications. The study shows NVIDIA Xavier NX and TX2 results in complex workloads due to their superior performance and efficiency [13–15]. Furthermore, research suggests that embedded GPUs are suitable for infrared detector algorithms on board [8].

Works presented by Steenari *et al*. and Marques *et al*.

---

[*]Corresponding author. E-Mail: luis.garces@uni.lu

analyze high-performance processors and FPGAs for onboard processing and ML applications. The challenges include the gap between the processing power of COTS devices and radiation-hardened options and the longer support times required for COTS, which can impact the mission reliability [6, 16].

In this context, this paper introduces a platform for testing artificial intelligence and machine learning in satellite communication payloads utilizing readily available AI-capable consumer chipsets. The Artificial Intelligence Satellite Telecommunications Testbed (AISTT) is the development of the ESA project Satellite Signal Processing Techniques using a Commercial Off-The-Shelf AI Chipset (SPAICE) [17]. It also examines the selection of hardware and its application, along with a detailed explanation of the ML payload design. Furthermore, the outcomes of the AI/ML onboard payload integration are explained.

## 2 Testbed description

The use of a partially regenerative satellite payload, implemented as a low-PHY layer onboard, motivates the use of flexible payload and beam management algorithms accelerated with AI on the satellite. Regenerative payloads can handle inter-satellite links to relay connectivity to multiple gateways, increasing the link budget at the user link and spectral efficiency at the feeder link and simplifying the implementation of user and gateway handover. The chosen application for the SPAICE project involves a versatile software-controlled satellite payload linked to a multibeam Direct Radiating Array (DRA) antenna with hybrid beamforming capabilities where the payload adjusts the bandwidth, power, and width of the beams [18, 19].

In general, integrating ML techniques into the onboard payload involves training and validating the payload model offline with a set of input/output values, as depicted in Figure 1. This model is then incorporated into the satellite payload processor to carry out the inference process. The feeder uplink signals are processed onboard depending on the demand requirements for the satellite's coverage area. This will adapt the signals based on the model, generating control instructions for the low physical layer, improving transmission efficiency, managing network congestion, and optimizing bandwidth and power.

### 2.1 Mission scenario

The mission scenario imposes constraints on the AISTT due to the type of signal, bandwidth, and power required for data communication. The refer-ence mission scenario includes the onboard payload design for 12U CubeSat providing coverage to Europe in LEO sun-synchronous orbit (SSO), featuring a high-duty cycle (>50%) and maximum power output (<100 W). The design specifically considers an altitude of 600 km and coverage of seven beams [19], although only two user equipment located on two different beams are emulated for testbed simplification. These payload requirements can be escalated to other mission scenarios.

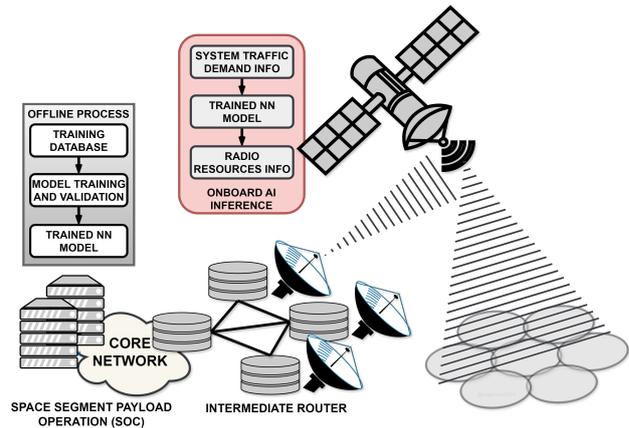

**Figure 1:** Onboard AI Payload application.

### 2.2 AI onboard payload architecture

AI onboard payload includes the inference system, the software-defined radio (SDR) RF front-end, the firmware, and the configuration software, forming a platform capable of assessing and enhancing the mission payload.

Identifying a device with sufficient computing power, appropriate energy usage, and the ability to meet the standards for onboard and standalone applications is crucial. Studies published in [9] and [10] revealed that the AMD/XILINX Versal ACAP AI Core and Edge series are good candidates for the implementation of AI algorithms for standalone applications with strong constraints on size, power consumption and sufficient performance per watt. Unfortunately, no Versal ACAP with radio frequency capabilities is on the market.

Based on the evaluation of existing COTS, a feasible implementation separates the inference process and radio frequency operation into two different chipsets, interconnecting them with a high-speed serial interface. This approach enables using an AI-capable chipset in conjunction with an SDR, thus isolating the inference procedure from the low physical layer, and the RF front-end. It offers greater versatility and performance and provides a comprehensive solution that combines powerful AI processing capabilities with readily available RF interfaces.

Taking into account the available Versal AI COTS boards, two exponents from different families have been summarized in Table 1. The AMD VCK190 Evaluation Kit is powered by a Versal AI Core chip with a dual-core Computing Processing Unit (CPU), a dual-core Real-Time Processing Unit (RTPU), 400 AI Engines (AIE), 1968 Digital Signal Processing Engines (DSPE), and 1968 k Logic Blocks (CLB). It can execute floating point operations up to 8 TFLOPs with a maximum efficiency of 91.9 GFLOPs/W. On the other hand, the iWave Systems iW-RainboW-G57D Development Kit, which hosts a Versal AI Edge SoC with 34 AI Engines ML (AIE-ML), 324 DPSE, and 329 k CLBs, is designed for edge processing with lower computer capacity (up to 1.9 TFLOPs) but better performance (up to 95 GFLOPs/W). This is achieved thanks to the chip's reduced power consumption (around 20 W) and the optimized AI Engines for machine learning operations.

| COTS AI/ML | VCK190 | iW-G57D |
|---|---|---|
| AI Chip | VC1902 | VE2302 |
| Family | ACAP AI Core | ACAP AI Edge |
| CPUs | Cortex A72×2 | Cortex A72×2 |
| RTPUs | Cortex R5F2×2 | Cortex R5F2×2 |
| AIEs | 400 | 34 (AIE-ML) |
| DSPEs | 1968 | 324 |
| CLBs | 1968 $k$ | 329 $k$ |
| Chip power | $\approx 87\ W$ | $\approx 20\ W$ |
| CC INT8 OPs/W | $13.6 - 133\ T$ $156G - 1.53\ T$ | $3.2 - 23\ T$ $160G - 1.15\ T$ |
| CC FP32 FLOPs/W | $3.2 - 8\ T$ $36.8 - 91.9\ G$ | $0.7 - 1.9\ T$ $35 - 95\ G$ |
| Type | Dev. Kit | SoM + Carrier |
| Expansion | FMC+, SFP+ | FMC+ |
| Power supply | 180 $W$ | 60 $W$ |
| Board Size | $24 \times 19\ cm^2$ | $12 \times 12\ cm^2$ |

Table 1: Considered AI/ML-Capable COTS boards.

The power supply and size of the board are constraints that define the selection of the development kit. The proposed reference scenario makes the AMD VCK190 an infeasible solution.

A similar analysis has been performed for the SDR. The HiTech Global ZRF-FMC-4A4D, with an FMC+ VITA 54.7 interface, has been selected. This small form-factor addon board powers a third-generation AMD radio frequency SoC (RFSoC) with four 14-bit RF ADCs and DACs, consuming up to 45 W.

Figure 2 shows the onboard satellite payload firmware, highlighting the ML models implemented using Versal AI Engines. The payload receives the traffic demand ($R$), beam pointing angles ($Az$, $El$), and beamforming coefficients phasors ($e^{j\theta W}$).

The *Flexible Payload Conf.* module, a trained ML model, processes traffic demand and outputs the bandwidth ($BW$) for the uplink feeder signal ($fULnk$), as well as the equivalent isotropically radiated power ($EIRP$) per beam to the *Adaptive Beamformer* module, which is also an ML model that uses the beam pointing angles in addition to the $EIRP$. The inference process generates the coefficients module vector ($|W|$), which, with the coefficients phasors, is modified to obtain the beamforming coefficients ($W$) on the *Coef. gen* module.

In the *low PHY* layer, the two uplink modulated feeder signals ($fULnk_{chx}$) are processed using cyclic prefix orthogonal frequency division multiplexing (*CP-OFDM*) by fast Fourier transformation (*FFT*). Subsequently, the time domain beamforming is performed by applying the channel matrix to the signal vector multiplication by the embedded multipliers in the RFSoC (DSP48E2).

The digital front-end interfaces the digital and analog RF signals, applying frequency division multiplexing (FDM) and digital up-conversion (DUC) to the 64 antenna signals before transforming them into four analog downlink RF signals ($IFDLnk$). At the same time, the two analog RF channel signals emulating the return link ($IFULnk$) are digitized and sent back to the signal source without processing. For simplicity, the low PHY layer in the return link is omitted.

### 2.3 AISTT integration

The Artificial Intelligence Satellite Telecommunication Testbed (AISTT) architecture integrates the payload, scenario generation, base station emulator, channel emulation, and user equipment to create a comprehensive platform for evaluating and optimizing the mission payload performance. Figure 3 depicts the AI/ML onboard payload integration with the other parts of the auxiliary subsystems.

The *Scenario Generator* is a MATLAB script within the payload control center. It produces the inputs for the ML algorithm, including traffic demand ($R$), beam pointing angles ($Az$, $El$), and beamforming coefficient phasors ($e^{j\theta W}$) every second. New traffic demand is generated every 37 s depending on the time of day, population density, and air and maritime traffic.

The *Base Station Generators* are two Next Generation NodeB distributed units (gNB-DU) running OpenAir-

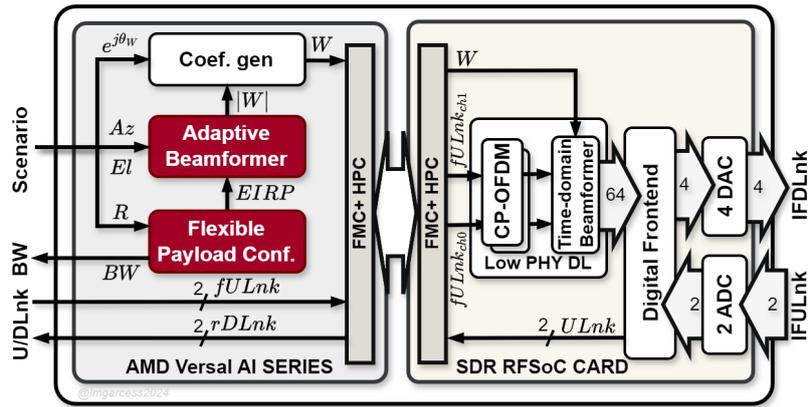

**Figure 2:** Onboard Payload Firmware Diagram.

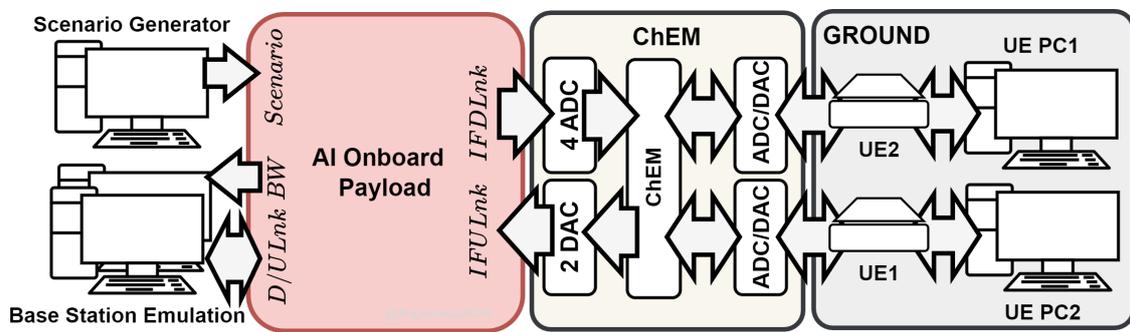

**Figure 3:** AISTT Functional Diagram.

Interface (OAI) that generate the feeder link signals and receive the return links. For implementation purposes, the partially regenerative functions running on the satellite payload are the ones on the feeder link low PHY layer, maintaining the rest of the functionalities on the OAI side. The MAC scheduler on the OAI adapts the beam beamwidth based on the results of the flexible payload.

On the other hand, the channel emulator (*ChEM*) receives the RF signal from the payload, applying adjustable channel effects such as delay, Doppler, and noise, and regenerates them for the user's equipment. It receives the signals from the 64 antennas of the satellite DRA [19] downlink signal multiplexed through four RF connectors that are demultiplexed in the *ChEM* and converted to two beams using channel matrix multiplication with the information of the beams of interest.

In the final stage, two user equipment (*UE*) capture the two beams from the *ChEM*. Such *UEs* are placed in different satellite beams, aggregating all the traffic demand corresponding to all the users served by each beam. At the same time, the OAI UE generates user information that will be retransmitted to the gNB-DU (*ULnk*) to modify downlink requirements.

## 3 Preliminary results

Figure 4 shows the coverage zone, focusing on a part of Europe of a CubeSat with seven-beam coverage. The alignment of these beams depends on the satellite's orbital location. Additionally, the figure underscores the seven beams provided for the current time step and the particular orbital transit of the CubeSat relevant to our analysis.

Table 2 presents the performance results achieved by the payload for the reference LEO scenario. The resulting values exceed the SPAICE performance requirements, such as the signal-to-interference-plus-noise ratio (SINR), the average spectral efficiency, and the throughput. The demand match is quantified by the normalized mean square error (NMSE) of the resulting inference demand and the input capacity, considering the scenario's reference capacity. The maximum energy/power usage, the complexity of implementation, and the rapid response to modifications are presented in relation to both ML-based and non-ML-based solutions, showing favorable outcomes. The implementation complexity evaluates how fast the ML-based solution is compared to the non-ML-based solution, while the response time to modifications is presented as the speed increment percentage.

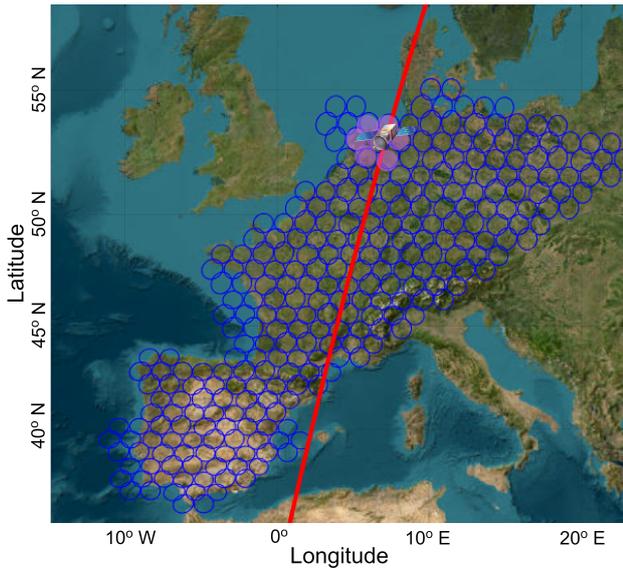

**Figure 4:** Functional Diagram.

| Description | Req. | AI/ML Payload |
|---|---|---|
| SINR [dB] | >6 | 7 |
| Aver. SE [b/s/Hz] | >1.5 | 2.1 |
| Throughput [Mbps] | >18 | 53.6 |
| Demand Match. | <0.4 | 0.07 |
| Max. Pwr. | <40% | 6.3% |
| Imp. Compl. [sec] | <60 | 3.3 |
| Response time | >90% | >97.58% |

**Table 2:** SPAICE performance requirements and implementations results.

## 4  Conclusions

The AISTT is a useful tool for evaluating various onboard payload scenarios and the effectiveness of ML techniques in satellite communication systems. With the use of COTS AI chipsets, the testbed offers a flexible and cost-effective solution to improve payload management and performance compared to space-qualified devices. The results show significant improvements in signal quality, spectral efficiency, and throughput, underscoring the potential for the integration of AI in space applications. The AISTT's ability to simulate different mission scenarios and hardware configurations provides valuable insights, ensuring that future satellite missions can be optimized for performance and efficiency.

## 5  Acknowledgment

This work has been supported by the European Space Agency (ESA), which funded it under Contract No. 4000134522/21/NL/FGL, "Satellite Signal Processing Techniques using a Commercial Off-The-Shelf AI Chipset (SPAICE)." Please note that the views of the authors of this paper do not necessarily reflect the views of ESA